# The Correlation Function of Rich Clusters of Galaxies in CDM-like Models

R. A. C. Croft and G. Efstathiou
*Department of Physics, University of Oxford, Keble Road, Oxford, OX1 3RH. UK.*

**ABSTRACT**
We use ensembles of high-resolution CDM simulations to investigate the shape and amplitude of the two point correlation function of rich clusters. The standard scale-invariant CDM model with $\Omega = 1$ provides a poor description of the clustering measured from the APM rich cluster redshift survey, which is better fitted by models with more power at large scales. The amplitudes of the rich cluster correlation functions measured from our models depend weakly on cluster richness. Analytic calculations of the clustering of peaks in a Gaussian density field overestimate the amplitude of the N-body cluster correlation functions, but reproduce qualitatively the weak trend with cluster richness. Our results suggest that the high amplitude measured for the correlation function of richness class $R \geq 2$ Abell clusters is either an artefact arising from incompleteness in the Abell catalogue, or an indication that the density perturbations in the early universe were very non-Gaussian.

## 1 INTRODUCTION

Investigations of the spatial distribution of rich clusters provided some of the earliest evidence for large scale irregularities in the Universe (Kiang & Saslaw 1969, Yu & Peebles 1969, Bogart & Wagoner 1973, Hauser & Peebles 1973). More recently, various redshift surveys of clusters selected from the Abell catalogue (Abell 1958) have been used to estimate the rich cluster two-point correlation function $\xi_{cc}(r)$ (*e.g.* Bahcall & Soneira 1983, Klypin & Kopylov 1983, Postman, Huchra & Geller 1992, hereafter PHG). Two new cluster catalogues have been constructed from digitised scans of photographic plates, and redshift surveys have been carried out by Dalton et al. (1992) (the APM Cluster Catalogue), hereafter DEMS and Nichol et al. (1992) (the Edinburgh-Durham Cluster Catalogue, or EDCC). New redshift surveys of X-ray selected samples of rich clusters are in progress.

It has been known since the work of Hauser and Peebles (1973) that rich clusters of galaxies are more strongly clustered than galaxies, but the amplitude of the cluster two point correlation function, and its dependence on cluster richness are controversial. Taken at face value, the three-dimensional clustering of rich clusters clusters is consistent with a scaling relation

$$r_0 = 0.4 d_c, \qquad \xi_{cc}(r) \approx 0.2(r/d_c)^{-1.8}. \qquad (1)$$

where $\xi_{cc}(r_0) = 1$ and $d_c$ is the mean intercluster separation related to the mean space density $n_c$ by $d_c = n_c^{-1/3}$. Bahcall & Burgett (1986), Bahcall & West (1992), and Bahcall & Cen (1992) (hereafter BC92) argue that equation (1) applies for clusters over the range $d_c = 30\ h^{-1}$Mpc * to $d_c \approx 90\ h^{-1}$Mpc.

Redshift surveys based on the APM Cluster Catalogue and EDCC have provided accurate estimates of the correlation length $r_0$ for clusters of richnesses intermediate to those of $R = 0$ and $R = 1$ Abell clusters. These surveys indicate $r_0 \approx 13-16\ h^{-1}$Mpc for clusters with $d_c \approx 35-46\ h^{-1}$Mpc, consistent with equation (1). However, these surveys are too small to provide accurate measurements on the clustering of richer clusters. The extrapolation of equation (1) to richer clusters relies on redshift surveys based on the the Abell catalogue (Abell 1958) and its southern counterpart (Abell, Corwin & Olowin 1989). For example, Bahcall and Soneira (1983) find $r_0 \approx 25\ h^{-1}$Mpc for Abell $R \geq 1$ clusters ($d_c \approx 55\ h^{-1}$Mpc), while Peacock & West (1992) find $r_0 = 21.1 \pm 1.3\ h^{-1}$Mpc for $R \geq 1$ Abell clusters and $r_0 = 45 \pm 5\ h^{-1}$Mpc for $R \geq 2$ Abell clusters ($d_c \approx 84\ h^{-1}$Mpc).

However, it is unclear whether these results for Abell clusters are correct. Various authors have presented evidence that the Abell catalogues, which were constructed by scanning photographic plates by eye, are affected by incompleteness on the plane of the sky which enhance the clustering amplitude measured in three dimensions (Sutherland 1988; Soltan 1988; Sutherland & Efstathiou 1991; Efstathiou *et al.* 1992, hereafter EDSM). A particularly strong case for patchy incompleteness in the Abell $R \geq 0$ catalogue has been presented by EDSM, who compare the machine based APM survey with the PHG redshift survey of Abell clusters. The clusters in these surveys have comparable space densities, but the redshift-space correlation function of the APM sample is isotropic on large scales with a low amplitude ($r_0 \approx 13\ h^{-1}$Mpc), while the correlation function for the Abell clusters is highly anisotropic and has a much higher amplitude. The effect of patchy incompleteness on samples of richer Abell clusters is more controversial. Sutherland (1988), EDSM and others have argued that the correlation functions of the $R \geq 1$ Abell catalogue may have been significantly overestimated whereas some authors, *e.g.* Bahcall and West (1992) and Peacock & West (1992), suggest that any biases are small.

---

* Throughout this paper we write the Hubble constant as $H_0 = 100 h$ km s$^{-1}$ Mpc$^{-1}$.

Equation (1) is therefore approximately correct for clusters with $d_c \approx 35\ h^{-1}$Mpc and agrees with results from machine measured surveys. However the extrapolation of equation (1) to higher richnesses relies exclusively on empirical results from the Abell catalogue. Since there is evidence for patchy incompleteness in the Abell sample, the richness dependence implied by equation (1) must be viewed with some skepticism. In this paper we give high weight to the machine based surveys of DEMS and Nichol *et al* (1992), since these are known to be free of large redshift-space anisotropies of the clustering pattern, and we give low weight to the results from the Abell catalogue for the reasons outlined in the previous paragraph. We refer the reader to the papers by Dalton *et al* (1992) and Efstathiou *et al* (1992) for more detailed discussions of the observations.

In this paper, we use N-body simulations to investigate the clustering of rich clusters from Gaussian initial conditions. Our aim is to determine the shapes of the cluster correlations and how these depend on the amplitude of the mass fluctuations and on cluster richness. A number of authors have investigated these problems theoretically using the statistical properties of Gaussian density fields (see *e.g.* Kaiser 1984; Bardeen *et al.* 1986, hereafter BBKS; Bardeen, Bond & Efstathiou 1987; Holtzman & Primack 1993; Mann, Heavens & Peacock 1993), but it is not clear how well these analyses compare with the full non-linear problem. White *et al.* (1987) and Bahcall & Cen (1992) have determined the two-point cluster correlations from N-body simulations of CDM-like universes, but our computations sample a larger volume of space than these studies, and have higher spatial resolution than the particle-mesh simulations of Bahcall & Cen.

The layout of this paper is as follows. The numerical simulations and the cluster selection algorithm are described in Section 2. The cluster correlations are discussed in Section 3 and are compared to an analytic model based on the statistical properties of Gaussian density fields in Section 4. Section 5 compares the APM cluster redshift survey (DEMS) with simulated APM cluster catalogues generated from the N-body models. Our conclusions are summarized in Section 6.

## 2 NUMERICAL SIMULATIONS AND CLUSTER SELECTION

### 2.1 Numerical simulations

We carried out 30 simulations of CDM-like universes in cubical volumes of comoving box length $\ell_B = 300\ h^{-1}$Mpc using a particle-particle-particle-mesh (P$^3$M) code ( Efstathiou & Eastwood 1981, Efstathiou *et al.* 1985). Each simulation followed the non-linear gravitational evolution of $10^6$ particles in a $256^3$ mesh with a spatial resolution of $\simeq 80 h^{-1}$kpc. The high resolution of our simulations ensures that we can follow the formation and evolution of mass concentrations that can be identified with rich clusters of galaxies (see Section 2.2).

The initial power spectrum of our models is given by:

$$P(k) \propto \frac{k}{[1 + (ak + (bk)^{3/2} + (ck)^2)^\nu]^{2/\nu}} \qquad (2)$$

(where $\nu = 1.13$, $a = 6.4/\Gamma\ h^{-1}$Mpc, $b = 3.0/\Gamma\ h^{-1}$Mpc, $c = 1.7/\Gamma\ h^{-1}$Mpc). This power spectrum applies for scale-invariant CDM models with low baryon densities $\Omega_B \ll \Omega_0$ and $\Gamma = \Omega_0 h$ (Bond & Efstathiou 1984). Thus the 'standard' CDM model with $\Omega_0 = 1$ and $h \sim 0.5$ (Davis *et al.* 1985) has $\Gamma \approx 0.5$. Ensemble A consists of 10 simulations of the standard CDM model with $\Gamma = 0.5$. There is considerable evidence that large-scale clustering of galaxies and clusters is better described by a power spectrum with $\Gamma \approx 0.2$ (Efstathiou, Sutherland & Maddox 1990, Efstathiou 1993). We have therefore run two additional ensembles each consisting of 10 simulations of spatially flat models with $\Gamma = 0.2$, using the same phases as the models in ensemble A to generate the initial particle displacements. Ensemble B has a cosmological constant, $\lambda = \Lambda/(3H_0^2) = (1 - \Omega_0) = 0.8$, and ensemble C has $\Gamma = 0.2$ and $\Omega_0 = 1$.

The parameters of the models are given in Table 1. $A_i$ is the amplitude of the initial linear fluctuations relative to the white-noise level at the Nyquist frequency of the particle grid used in setting up the initial conditions (Efstathiou *et al.* 1985). The column labelled $(\sigma_8)_i$ gives the initial *rms* amplitude of the linear mass fluctuations in spheres of radius 8 $h^{-1}$Mpc. Each model was evolved until the value of $\sigma_8$ was equal to 1.0 according to linear theory. The expansion factor required to reach this value of $\sigma_8$ from the start of the simulations is given by $a_{final}$.

**Table 1.** Parameters for the CDM-like models under investigation

| Ensemble | $\Gamma$ | h | $\lambda$ | $A_i$ | $(\sigma_8)_i$ | $a_{final}$ |
|---|---|---|---|---|---|---|
| A | 0.5 | 0.5 | 0.0 | 0.70 | 0.236 | 4.24 |
| B | 0.2 | 1.0 | 0.8 | 0.64 | 0.311 | 4.50 |
| C | 0.2 | 1.0 | 0.0 | 0.40 | 0.194 | 5.15 |

Each simulation took about two days on a Decstation 5000/240 with 168Mbytes of memory. With the large mesh used here, the calculations are mesh dominated; towards the end of the calculations, when the particle distributions are clustered most strongly, computing the short-range forces takes only about 25% of the time required to compute the mesh forces.

## 2.2 Cluster selection

To select candidate centres for clusters we found groups of particles linked together by less than 0.1 times the mean interparticle separation using the percolation algorithm described by Davis *et al.* (1985). We then searched for clusters by counting particles within radius $r_c$ of the percolation centres. We calculated the centre of mass of each cluster and deleted any cluster within $r_c$ of a cluster of a larger mass. We then repeated the cluster finding by counting particles within $r_c$ of the centre of mass of each cluster, recomputing the centre of masses and deleting overlapping clusters. This step was repeated twice to determine the final list of clusters. This algorithm is the same as that used by White, Efstathiou & Frenk (1993). Having generated a cluster catalogue from a simulation for a specified value of $r_c$, we order the clusters by mass. By applying a lower mass limit, we can generate samples of different space densities $n_c$ and mean intercluster distance $d_c = n_c^{-1/3}$.

This algorithm, which finds mass concentrations within spherical volumes in three-dimensions, is similar to the algorithms used to identify clusters from two-dimensional catalogues of galaxies. Abell's (1958) algorithm is designed to find clusters of galaxies within a metric radius of $R_A = 1.5 \ h^{-1}$Mpc. Clusters of galaxies in the APM survey are found using an Abell-like algorithm with a smaller metric radius of $0.75 \ h^{-1}$Mpc, but since galaxies closer to the cluster centre are given higher weight than those in the outer parts, the effective cluster radius in the APM sample is slightly smaller. We have generated cluster catalogues using $r_c = 1.5 \ h^{-1}$Mpc and $0.5 \ h^{-1}$Mpc. As we will show in the next section, our results are almost independent of the value of $r_c$.

The simple prescription that we have adopted avoids having to assign galaxies to mass points, which would introduce uncertain parameters and further assumptions since the galaxy formation process is so poorly understood. For a given choice of cluster radius $r_c$ our procedure results in a unique list of clusters ordered by mass. Provided, therefore, that there is an approximate monotonic relation between the masses of rich clusters and their luminosity, our prescription should provide an accurate match to observed samples limited by cluster richness, even though we make no specific model for assigning galaxies to the mass. This is an important simplification which we feel is reasonable and necessary to make the problem tractable. One can imagine situations where our assumptions would be incorrect, for example if cluster luminosities were modulated by some correlated physical mechanism in analogy with models of 'cooperative galaxy formation' as discussed, for example by Bower *et al.* (1993). However, such mechanisms are so poorly understood that they have generally been ignored when comparing theoretical models to observations.

Figure 1 shows clusters in slices of depth $75 \ h^{-1}$Mpc at the final output times ($\sigma_8 = 1$) from a model from ensemble A and one with identical initial random phases from ensemble B. The open circles show clusters identified within $r_c = 1.5 \ h^{-1}$Mpc and the filled points show clusters identified within $r_c = 0.5 \ h^{-1}$Mpc. These catalogues have $d_c = 35 \ h^{-1}$Mpc, comparable to the mean intercluster distance of $\mathcal{R} \geq 20$ clusters in the APM survey (EDSM). The two pictures are visually extremely similar despite the differences in the power-spectra. There is also a good correspondence between the clusters found within the two search radii. If all clusters had identical density profiles, our cluster catalogues would be independent of $r_c$. In fact, this is nearly the case and the clusters found with $r_c = 0.5\ h^{-1}\mathrm{Mpc}$ have approximately one third of the mass of clusters identified with $r_c = 1.5\ h^{-1}\mathrm{Mpc}$, as expected for clusters with near isothermal profiles $M(r) \propto r$ (see also Figure 1 of White et al. (1993)).

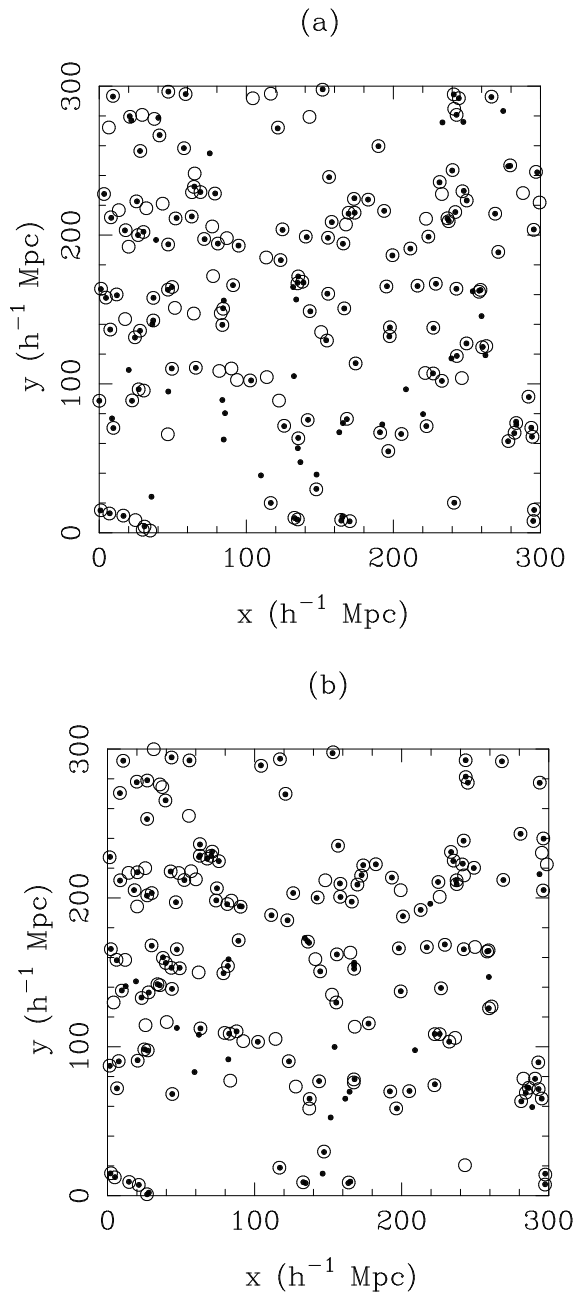

**Figure 1.** A 75 $h^{-1}$Mpc slice through simulations with identical initial random phases picked from (a) ensemble A, and (b) ensemble B. Clusters are selected with $r_c = 0.5 \ h^{-1}$Mpc (dots) and $r_c = 1.5 \ h^{-1}$Mpc (Open circles), and with $d_c$ (mean intercluster separation) = 35 $h^{-1}$Mpc. The symbol sizes are not to scale.

## 3  THE TWO-POINT CLUSTER CORRELATION FUNCTION

For each cluster catalogue we compute the two-point correlation function using a direct estimator:

$$\xi_{cc}(r) = \frac{N_p}{n_c^2 V\, dV} - 1, \qquad (3)$$

where $N_p$ is the number of cluster pairs in the radial bin of volume $dV$ centred at $r$, $n_c$ is the mean space density of the cluster catalogue, $V$ is the volume of the simulation. We use all clusters in a simulation taking advantage of the periodic boundary conditions.

For each simulation we compute $\xi_{cc}$ as a function of cluster richness and epoch. In the next subsection, we discuss the evolution of the correlation functions. We then investigate how $\xi_{cc}$ depends on the cluster selection radius $r_c$ and on cluster richness.

### 3.1  Evolution of the two-point correlation function

In Figure 2, we pick subsamples of clusters with a mean separation $d_c = 35\ h^{-1}\mathrm{Mpc}$, close to that of APM $\mathcal{R} \geq 20$ clusters (cf. Figure 1). We plot the correlation functions at various output times with the values of $\sigma_8$ given in the Figure.

The amplitude of the cluster correlations is insensitive to the amplitude of fluctuations in the density field. This is expected because these clusters are much more highly clustered than the mass fluctuations, even when the latter achieve values of $\sigma_8 \approx 1$. The insensitivity of the clustering on $\sigma_8$ is observed in our simulations even for poor clusters with $d_c \sim 10\ h^{-1}\mathrm{Mpc}$. There has been some controversy over the amplitude of the mass fluctuations in the universe at the present epoch (see for example Couchman & Carlberg (1992), who propose an $\Omega = 1$ CDM universe with $\sigma_8 = 1$). However, a recent analysis of the abundances and masses of rich clusters of galaxies gives

$$\sigma_8 = (0.57 \pm 0.05)\, \Omega_0^{-0.56} \qquad (4)$$

(White, Efstathiou & Frenk 1993) and in the rest of this paper we use equation (4) to select appropriate output times in the simulations. Note that $\sigma_8$ in equation (4) is the value extrapolated to the present day using linear pertubation theory, and is extremely insensitive to the shape of the initial fluctuation spectrum. As Figure 2 shows, fortunately we do not have to specify $\sigma_8$ very precisely to make accurate predictions of the cluster correlations. For our low density ensemble, a best fit to the COBE data requires that $\sigma_8$ be slightly lower: $\sim 1.0$. The 10 models we plot results for therefore have this value at the present day. We have also run 4 additional low density models with $\sigma_8 = 1.34$, and find the results to be indistinguishable from the less dynamically evolved models.

In Figure 2 we have plotted the data points for the APM clusters computed by EDSM. We show points only out to $50\ h^{-1}\mathrm{Mpc}$ since the results at larger radii are consistent with zero (eg. compare figure (10) of EDSM with figure (1) of DEMS). Note further that we plot Poisson error bars on the data points which we argue in Section (5) underestimate the true errors by as much as 50%. Although the observed correlations were computed in redshift space, we show in Section 5 that the distortions between redshift space and real space in the simulations are small. The $\Gamma = 0.5$ curves lie below the APM data points at separations $\gtrsim 8\ h^{-1}\mathrm{Mpc}$. The two $\Gamma = 0.2$ models give much better agreement with the observations.

### 3.2  Changing the radius of cluster selection

Figure 3 shows the effect on each of our models of increasing the radius used to define clusters in the simulations from $r_c = 0.5\ h^{-1}\mathrm{Mpc}$ to $1.5\ h^{-1}\mathrm{Mpc}$. Since the correlation functions are so insensitive to the amplitude of the mass fluctuations (Figure 2), we plot results for one value of $\sigma_8$ for each ensemble. The cluster catalogues were limited in mass to give an intercluster separation of $d_c = 35\ h^{-1}\mathrm{Mpc}$.

As noted in Section 2, there is a good correspondence between cluster catalogues with the same value of $d_c$ but different values of $r_c$ (see Figure 1). Accordingly, the correlation functions for $r_c = 0.5$ and $1.5\ h^{-1}\mathrm{Mpc}$ shown in Figure 3 are very similar. The correlation functions for $r_c = 0.5\ h^{-1}\mathrm{Mpc}$ are systematically lower than those for $r_c = 1.5\ h^{-1}\mathrm{Mpc}$ by about 10–20%, but these differences are much smaller than the observational errors on estimates of $\xi_{cc}$ from real cluster catalogues. The independence of $\xi_{cc}$ on $r_c$ applies for all of the cluster catalogues we have generated, i.e. for $d_c$ in the range 10–70 $h^{-1}\mathrm{Mpc}$.

As explained in Section 2, we chose $r_c = 0.5\ h^{-1}\mathrm{Mpc}$ to model the selection of clusters in the APM survey, and $r_c = 1.5\ h^{-1}\mathrm{Mpc}$ to model the selection of Abell clusters. Figure 3 would lead us to expect that Abell and APM cluster catalogues with the same value of $d_c$ would have nearly the same amplitude of $\xi_{cc}$. We plot $\xi_{cc}$ for the APM $\mathcal{R} \geq 20$ clusters, as in Figure 2. The filled stars show results for the redshift survey of $R \geq 0$ Abell clusters (PHG) corrected for redshift-space anisotropies by EDSM. The Abell sample has $d_c = 38\ h^{-1}\mathrm{Mpc}$, which is nearly the same as the value for the APM sample. The results for the two catalogues are very similar, and are well described by the $\Gamma = 0.2$ models. There is a marginal tendency for the Abell data points to lie higher than those for the APM survey, and this is in the direction suggested by the N-body results.

As mentioned in the introduction, the clustering in the PHG sample agrees with that measured in the APM sample only after the Abell catalogue has been corrected for incompleteness on the plane of the sky. The two-point correlation function for the uncorrected $R \geq 0$ catalogue has about twice the amplitude of the APM correlation function (see PHG and EDSM). Clearly, in the class of models con-sidered here, such a large difference cannot be explained by differences in the cluster selection. EDSM show that the redshift space correlation function for the PHG sample is highly elongated along the line-of-sight, so that $\xi_{cc} \sim 1$ out to line-of-sight separations of $\sim 100~h^{-1}$Mpc for projected pair separations of $\lesssim 10~h^{-1}$Mpc. These large anisotropies provide strong evidence that the clustering pattern in the $R \geq 0$ Abell is enhanced by inhomogeneities in the completeness of the catalogue. The correction procedure described by EDSM removes these biases and brings $\xi_{cc}$ for the Abell sample into agreement with the results for the APM sample.

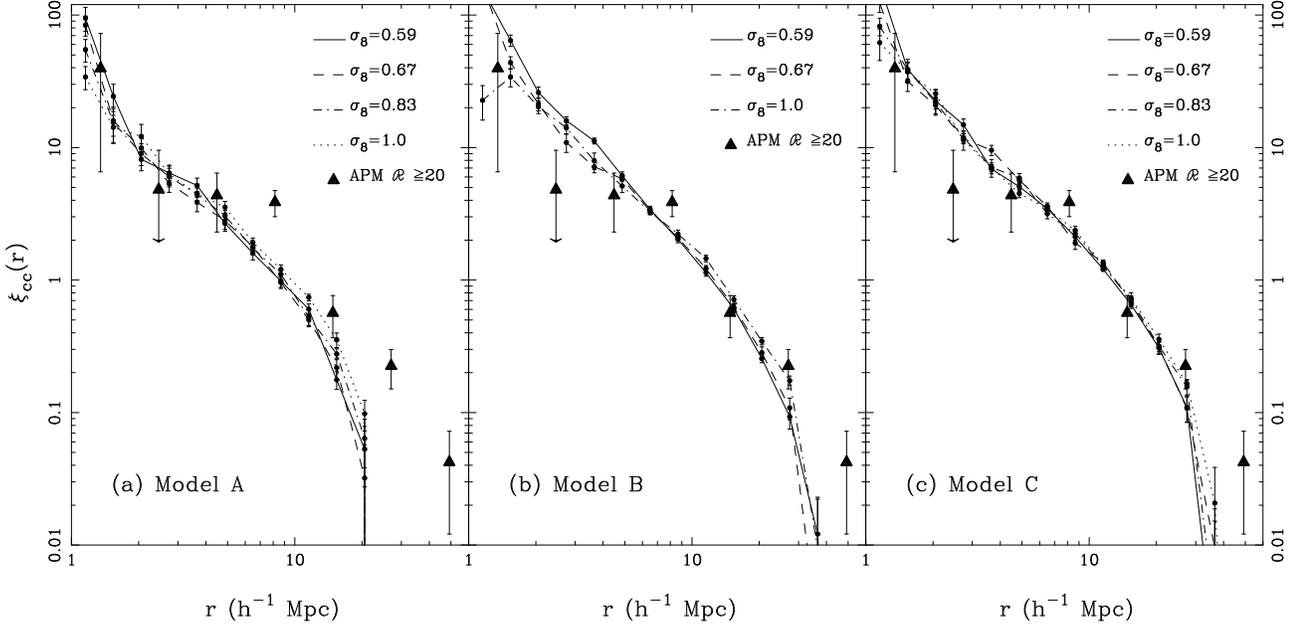

**Figure 2.** Time variation of the cluster correlation functions for clusters with $d_c = 35\ h^{-1}$ Mpc selected with a search radius of $r_c = 0.5\ h^{-1}$ Mpc. Figure 2(a) shows ensemble A, *i.e.* standard CDM with $\Omega_0 = 1$ and $h = 0.5$; (b) shows ensemble B in which $\Gamma = 0.2$, $\Omega_0 = 0.2$ and $\lambda = 0.8$; (c) shows ensemble C in which $\Gamma = 0.2$ and $\Omega_0 = 1$. The error bars on the simulation curves show one standard deviation on the mean and were computed from the scatter within each ensemble. The amplitudes of the mass fluctuations according to linear theory, $\sigma_8$, at the output times plotted are listed in each panel. The triangles show points for APM $\mathcal{R} \geq 20$ clusters from Figure 10 of EDSM.

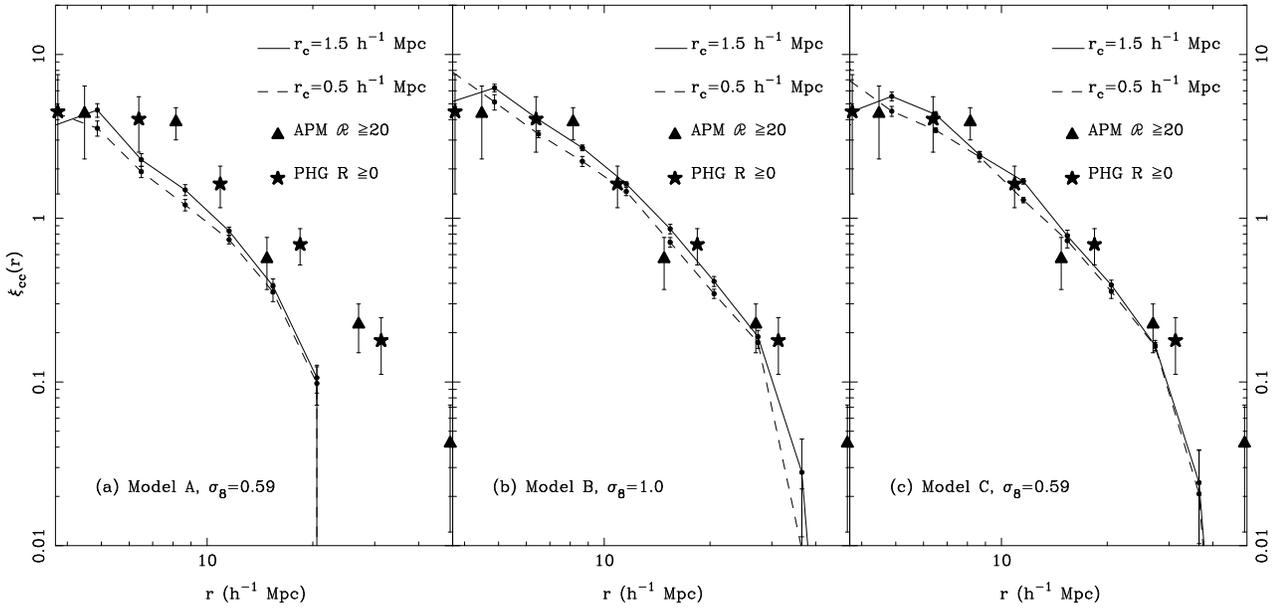

**Figure 3.** The dependence of the cluster correlation function on the cluster search radius, $r_c$, for the three ensembles at an amplitude $\sigma_8 = 0.59$ (ensembles A and C) and $\sigma_8 = 1.0$ (ensemble B). The cluster catalogues have $d_c = 35\ h^{-1}$ Mpc. We plot the APM data points, as in Figure 2. The points labelled PHG show the correlation function (corrected for projection biases) for the Postman, Huchra & Geller redshift survey of Abell clusters with $R \geq 0$ (from Figure 10 of EDSM).

In summary, for each ensemble we find that cluster catalogues with similar values of $d_c$ have similar two-point correlation functions. The clustering properties are therefore dependent on the shape of power-spectrum of the mass fluctuations, but not on the way in which clusters are identified. This agrees with observational results for APM and Abell clusters, but only if the Abell catalogue is corrected for biases in the cluster selection.

### 3.3 Richness dependence of the correlations

For each model, and for each of our two cluster selection radii, we generated 13 cluster catalogues at each output time by varying the lower richness bound so that the mean intercluster distance changed from $d_c \approx 10~h^{-1}$Mpc to $d_c \approx 70~h^{-1}$Mpc in steps of $5~h^{-1}$Mpc. We computed the two-point correlation function for these catalogues in logarithmic bins using the estimator of equation (3) and we computed the radial separation $r_0$ at which $\xi_{cc}(r_0) = 1$ by linear interpolation. The dependence of $r_0$ against $d_c$ provides a measure of how the amplitude of the correlation functions change with cluster richness.

In Figure 4, we plot the mean value of $r_0$ for each ensemble against $d_c$ for two values of $\sigma_8$ and for the two cluster selection radii. The error bars on $r_0$ show the $1\sigma$ errors determined from the scatter within each ensemble. As expected from the previous subsections, the amplitude of the cluster correlation function is insensitive to the amplitude of the mass fluctuations and to the cluster detection radius. The results for the two $\Gamma = 0.2$ models are virtually indistinguishable, but lie significantly higher than those for $\Gamma = 0.5$. For each ensemble, the value of $r_0$ becomes extremely insensitive to the richness of the cluster sample for $d_c \gtrsim 30~h^{-1}$Mpc. We show in section 4 that the dependence of $r_0$ against richness is in qualitative agreement with analytic calculations of the clustering of peaks in a Gaussian density field.

The solid lines in Figure 4 show equation (1), which Bahcall & Cen (1992) argue provides a good fit to the observed cluster correlations. Clearly the strong dependence of $r_0$ on $d_c$ of this relation provides a very poor match to the results from our simulations. BC92 have run N-body simulations of CDM models and seem to find a stronger dependence of $r_0$ on $d_c$ than we find from our models for similar initial power spectra. In fact, BC92 conclude that equation (1) can be reproduced in a CDM model with $\Gamma \approx 0.1$. We make the following remarks concerning these discrepancies.
[1] BC92 run a single particle-mesh simulation containing $250^3$ particles for three sets of cosmological parameters in a cubical box of side $\ell = 400~h^{-1}$Mpc. The force resolution of their simulations ($\gtrsim 800~h^{-1}$kpc) is considerably worse than ours, there are some differences in the cluster finding algorithms, and their mass resolution is better than ours by a factor of 6.6. Clearly these differences will lead to differences in the structure of the clusters that form in the simulations. However, the insensitivity of our results to the cluster selection radius described in Section 3.2 suggests that the amplitude of the correlation function is probably not sensitive to the resolution of the code. Whereas we have run 10 simulations in each ensemble, BC92 run only one model and so sample a volume of space smaller than ours by a factor of 4.2. The errors on their correlation functions will be larger than ours and it is plausible that the difference between our respective $r_0$–$d_c$ relations are caused by statistical fluctuations. In fact, over the range $10~h^{-1}\text{Mpc} \lesssim d_c \lesssim 40~h^{-1}\text{Mpc}$, where the errors in $r_0$ are small (Figure 4), our results agree well with BC92. BC92 use the same phases in each of their models (Bahcall, private communication) and this could explain why they see a stronger dependence of $r_0$ on $d_c$ in each of their models. From the trends shown in Figure 4, a $\Gamma = 0.1$ model should have slightly higher values of $r_0$ than a $\Gamma = 0.2$ model, but we would expect $r_0$ to be nearly independent of cluster richness.
[2] Equation (1) predicts $r_0 \approx 22~h^{-1}$Mpc for Abell $R \geq 1$ clusters and $r_0 \approx 38~h^{-1}$Mpc for Abell $R \geq 2$ clusters. As summarized in Section 1, redshift surveys of Abell clusters do indeed give values of $r_0$ compatible with these numbers, but there is considerable evidence that the Abell catalogue contains inhomogeneities that boost the amplitude of the clustering pattern. For example, Sutherland (1988) and Efstathiou et al. (1992) find $r_0 \approx 14~h^{-1}$Mpc for Abell $R \geq 1$ clusters after correction for large scale anisotropies. As we have discussed in the introduction, in our view, the empirical basis for equation (1) is weak.

In Figure 5 we show the correlation functions from our simulations for two richness cuts matching the space densities of $\mathcal{R} \geq 20$ ($d_c \approx 35~h^{-1}$Mpc) and $\mathcal{R} \geq 35$ ($d_c \approx 45~h^{-1}$Mpc) APM clusters. The observational points are from Figure 9 of EDSM. As expected from Figure 4, the correlation functions for the models are weakly dependent on $d_c$. The APM data are in excellent agreement with the two $\Gamma = 0.2$ models, but not with results for $\Gamma = 0.5$. The richness dependence of $\xi_{cc}$ for our models is shown in Figure 7 below for a wider range of $d_c$ and compared to calculations of the clustering of peaks in a Gaussian density field.

### 4 COMPARISON WITH THE STATISTICS OF GAUSSIAN DENSITY FIELDS

In this section, we compare our results with calculations of the cluster correlation function using the statistics of Gaussian density fields. This problem has been tackled by various authors (*e.g.* Kaiser 1984; BBKS; Bardeen, Bond & Efstathiou 1987). Recently, Mann et al. (1993) have applied an analytic technique described by Bond & Couchman (1988) to compute the evolution of the correlation function of peaks in a Gaussian density field assuming that the peaks move according to the Zeldovich (1970) approximation. Holtzman and Primack (1993) have used similar techniques to those described by Bardeen et al. (1987) to compute the cluster correlation function in mixed dark matter models.

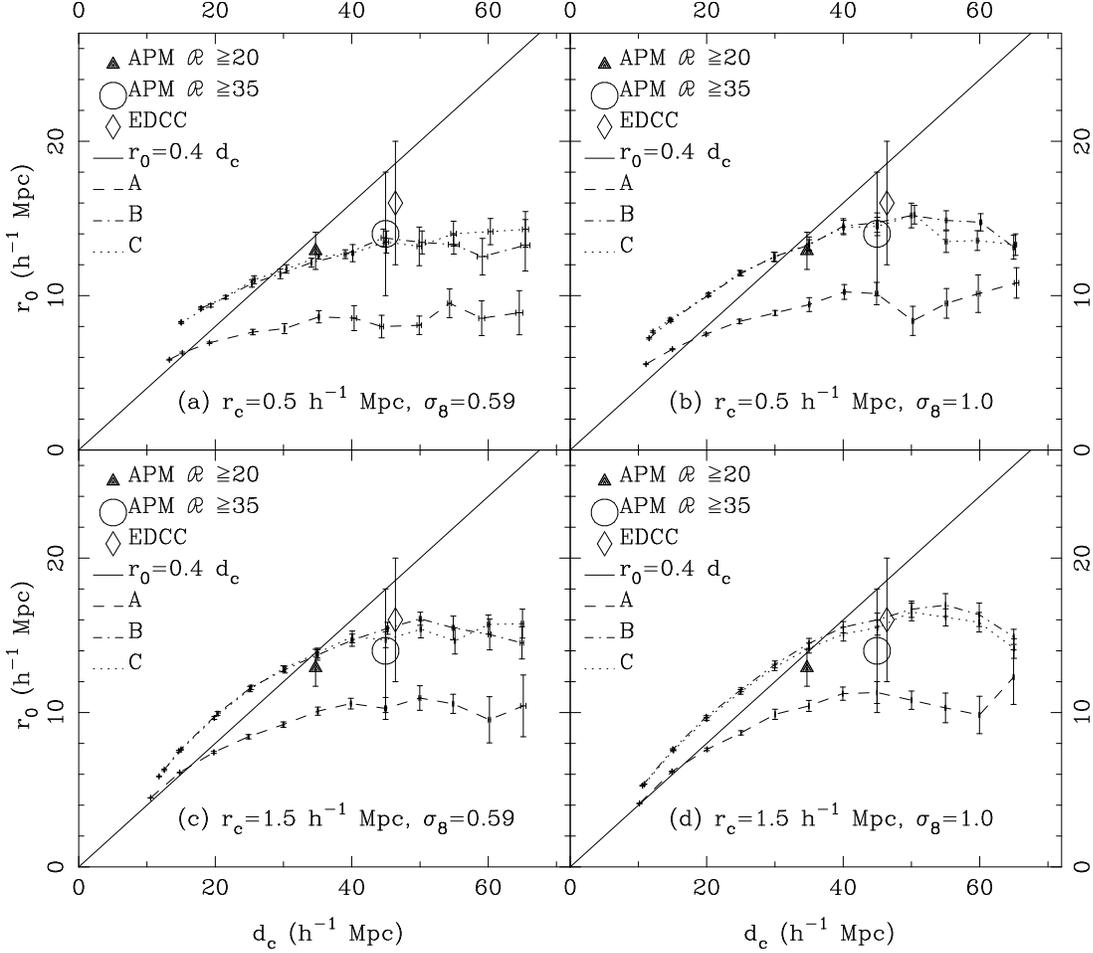

**Figure 4.** The dependence of $r_0$ for the cluster correlation function on intercluster separation, for two values of $\sigma_8$, 0.59 and 1.0, and two values of the cluster detection radius, $r_c = 0.5$ and $1.5\ h^{-1}$ Mpc. The solid lines show the relation $r_0 = 0.4 d_c$. We plot estimates of $r_0$ from the APM and EDCC redshift surveys.

We use the approximate formula for the peak-peak correlation function given in BBKS

$$\xi_{pk,pk} = \left(\frac{\langle \tilde{\nu} \rangle}{\sigma(R_s)} + 1\right)^2 \xi_{\rho,\rho} \qquad (5)$$

applied to peaks above a sharp threshold $\nu_t$. Here $\langle \tilde{\nu} \rangle$ is the averaged peak height above the threshold $\nu_t$ and $\sigma^2(R_s)$ is the variance of the density field smoothed by a Gaussian filter of width $R_s$. To apply equation (5) we need to select values for $R_s$ and $\nu_t$. Clearly, we should select peaks which can be identified with perturbations that will form clusters of galaxies; this suggests that we should choose $R_s$ so that the mass contained within the Gaussian window,

$$M(R_s) = (2\pi)^{3/2} \overline{\rho} R_s^3, \qquad (6)$$

where $\overline{\rho}$ is the mean mass density of the universe, is equal to the mass of a rich cluster of galaxies. Equation (6) gives

$$R_s = 2.8\ \Omega_0^{1/3} \left(\frac{M}{10^{14} M_\odot}\right)^{1/3} h^{-1} \text{Mpc}. \qquad (7)$$

The mass within an Abell radius, $r_A = 1.5\ h^{-1}$ Mpc (roughly the virialized region) of clusters with an abundance equal to half that of Abell $R \geq 1$ clusters ($d_c \simeq 66\ h^{-1}$Mpc($EDSM$)) is estimated to be about $4.2$–$5.5 \times 10^{14} h^{-1} M_\odot$ (White et al. 1993). The cluster mass functions of our N-body models are so steep that if we choose the amplitude of the mass fluctuations to approximately match the observed masses of clusters with $d_c = 63\ h^{-1}$ Mpc, then we find that the masses of clusters with the APM $\mathcal{R} \geq 20$ abundance ($d_c = 35\ h^{-1}$Mpc) are $\approx 2 \times 10^{14} h^{-1} M_\odot$ (see Figure 1 of White et al. al. (1993)). Thus, equation (6) gives about the correct masses for clusters over a wide range of abundances if we set $R_s \approx 5 \Omega_0^{1/3}\ h^{-1}$ Mpc.

Having chosen $R_s$, we can fix the threshold $\nu_t$ by matching the observed number density of clusters to the number density of peaks obtained by integrating the differential peak density distribution (BBKS equation 4.3)

$$\mathcal{N}_{pk}(\nu) d\nu = \frac{1}{(2\pi)^2 R_*^3} e^{\frac{-\nu^2}{2}} G(\gamma, \gamma\nu), \qquad (8)$$

where G is a function given in Appendix A of BBKS and the parameters $\gamma$ and $R_*$ are given by moments of the power spectrum.

This method of identifying peaks with real clusters is clearly very rough, so it is useful to compare the results with another method of choosing $R_s$ and $\nu_t$. Instead of fixing $R_s$ we choose $R_s$ and $\nu_t$ to reproduce the desired abundance of clusters and to satisfy

$$\nu_t \sigma(R_s) = \delta_c, \qquad (9)$$

where $\delta_c$, the critical density contrast for forming virialized objects by the present day, is set equal to 1.69 independent of $\Omega_0$ (see White et al., 1993). This scheme for choosing $R_s$ and $\nu_t$ has been used by BBKS and Mann et al. (1993).

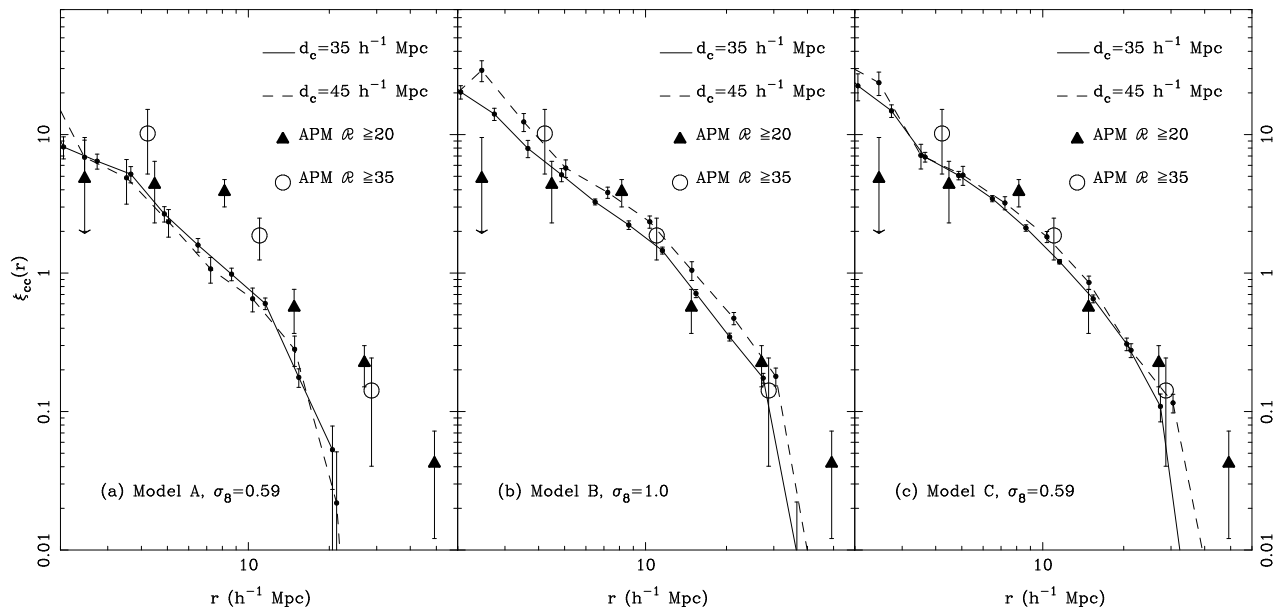

**Figure 5.** The correlation function for clusters identified with $r_c = 0.5\ h^{-1}\,\mathrm{Mpc}$ with same mean separation as APM clusters with richness $\mathcal{R} \geq 20$ ($d_c = 35\ h^{-1}\,\mathrm{Mpc}$) and $\mathcal{R} \geq 35$ ($d_c = 45\ h^{-1}\,\mathrm{Mpc}$).

Figure 6 shows calculations of the correlation length as a function of $d_c$. We have adopted $\sigma_8 = 0.59$ for the models A and C, and $\sigma_8 = 1$ for model B so that the figure can be compared with the results from the N-body simulations plotted in Figure 4. For the $\Omega = 1$ models, the two methods for choosing $R_s$ and $\nu_t$ give similar results at large values of $d_c$. However, for ensemble B ($\Omega_0 = 0.2$), equations (8) and (9) give $R_s$ closer to $5\ h^{-1}\,\mathrm{Mpc}$ at $d_c \gtrsim 40\ h^{-1}\,\mathrm{Mpc}$ than the value $R_s = 2.5\ h^{-1}\,\mathrm{Mpc}$ that we used to compute the dotted line. Comparing Figures 4 and 6 we see that the simple theory described here consistently *overestimates* the amplitude of the correlation function measured in the simulations. Mann et al. found a similar disagreement in comparing their theoretical predictions with the N-body results of BC92.

The discrepancies between Figure 6 and Figure 4 are worse for the higher value of $\sigma_8$, and are of the same order as the correction for the dynamical evolution of the mass-density fluctuations in equation (5) (approximated by the addition of $+1$ to $\tilde{\nu}/\sigma$). It might be thought that a more sophisticated treatment of the dynamical evolution of the correlations could improve the agreement. However, the computations of Mann et al. (1993), who use the Zeldovich approximation to model the motions of peaks, give a worse match to our results than the simple theory outlined above.

The dynamical contribution to the clustering is significant even in the simple analytic examples plotted in Figure 6 (see also Figure 7), i.e. $\tilde{\nu}/\sigma \sim 1$ and it is unclear how well the Zeldovich approximation describes the non-linear clustering in the numerical simulations. Furthermore, the peak prescription provides only a very crude guide to the locations and masses of clusters (see e.g. Bond et al. 1991). It is not particularly surprising therefore that the more elaborate calculations of Mann et al do not seem to provide a better description of the N-body results than the simple model described above. Evidently, a more precise description of the non-linear evolution of the mass fluctuations, possibly including cluster merging, is required to describe the N-body results. The *qualitative* trends shown in Figure 6 do match the N-body results; the relative amplitudes of $\xi_{cc}$ for the three models, and the levelling-off of $r_0$ with increasing $d_c$, are in rough agreement with the results described in Section 3.

In Figure 7, we plot the correlation functions for three cluster abundances at a single output time for each of the three ensembles. The dotted lines show the predictions of equation (5) with $R_s$ and $\nu_t$ fixed by equations (8) and (9). In each case the dotted lines overestimate the amplitude of the N-body curves. The shapes of the correlation functions for $\Gamma = 0.2$ are quite well described by the linear theory shape of $\xi_{\rho,\rho}$, whereas the N-body results for $\Gamma = 0.5$ are noticeably steeper than the linear correlation function on scales $\lesssim 10\ h^{-1}\,\mathrm{Mpc}$.

In summary, simple models based on the statistics of high peaks can provide a rough match to the clustering of rich clusters measured in our simulations. However, the amplitude of $\xi_{cc}$ is consistently overestimated by the high peak model.

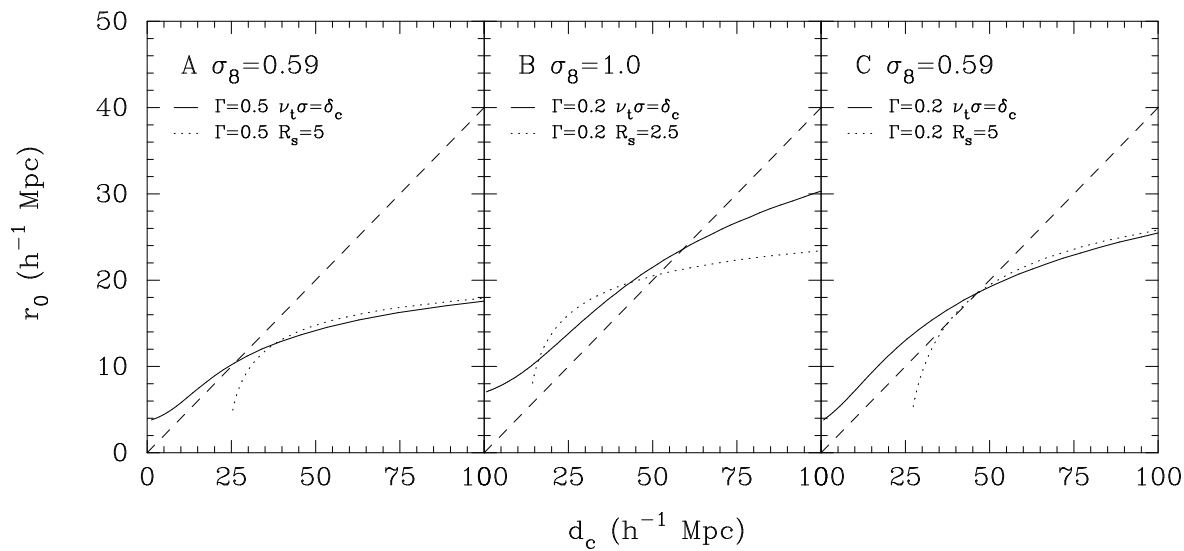

**Figure 6.** The dependence of the correlation length $r_0$ against intercluster distance $d_c$ computed from equation (5) for the three ensembles A-C. The solid lines show computations in which $R_s$ and $\nu_t$ satisfy equation (9) and the dotted lines show results for a fixed value of $R_s$ (listed in $h^{-1}$ Mpc in each panel). The dashed line shows the Bahcall & Cen relation (equation 1).

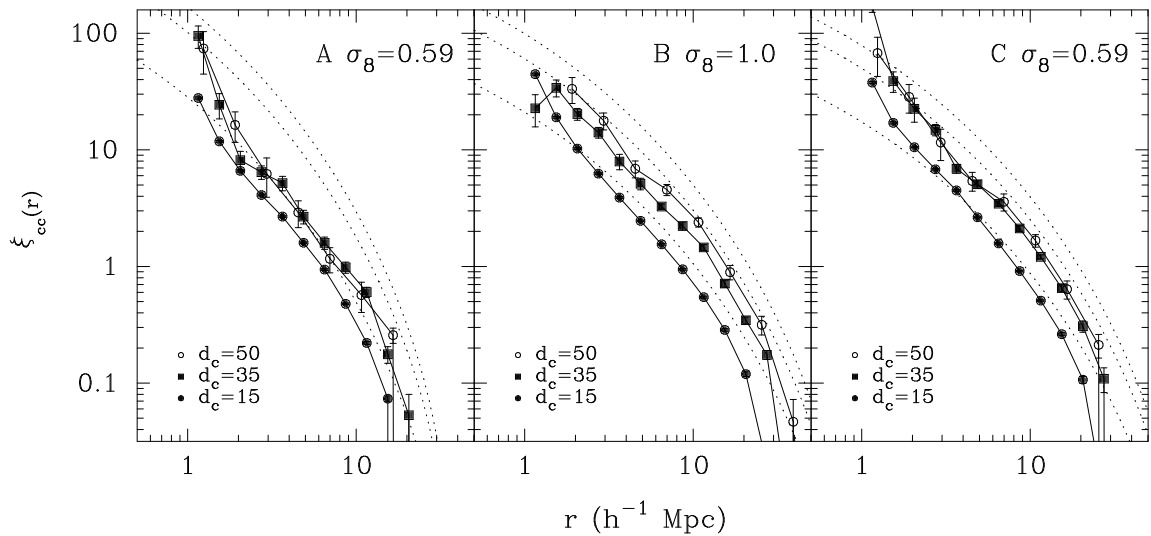

**Figure 7.** Cluster correlation functions for the three ensembles at $\sigma_8 = 0.59$ (ensembles A and C) and $\sigma_8 = 1.0$ for three values of $d_c$, $d_c = 15, 35$ and $50\ h^{-1}$ Mpc. The errors on the data points show one standard deviation of the mean. The dotted lines show the predictions of equation (5) in which $R_s$ and $\nu_t$ satisfy equation (9). Clusters in the N-body simulations were identified with $r_c = 0.5\ h^{-1}$ Mpc.

## 5 SIMULATED APM REDSHIFT SURVEYS

In this section, we use our simulations to generate cluster catalogues with the same selection criteria and volume as the APM cluster redshift survey of Dalton et al. (1992). We generate the simulated catalogues in the following manner:

We select rich cluster catalogues from the simulations normalised to the mean space density of the APM $\mathcal{R} \geq 20$ clusters and we replicate our computational box out to a redshift of 50000km s$^{-1}$ ( 5/3 boxlengths) using the periodic boundary conditions. After projecting the cluster positions onto the surface of a sphere, we discard all those outside the APM survey area. We select clusters at random using the selection function for APM clusters (see EDSM Figure 1) resulting in simulated catalogues with statistically similar number-redshift distributions and total number of clusters as the real survey. The two point cluster correlation function in redshift space for the simulated catalogues was computed by cross-correlating with a random catalogue and using the estimator of equation (1) from DEMS.

Our results are shown in Figure 8. The lines in each panel show $\xi_{cc}$ averaged over 10 simulations of the APM survey, and the error bars show the $1\sigma$ variation of the mean. As above, the $\Gamma = 0.2$ models provide an excellent match to the observations. The cluster random peculiar velocities in the simulations average $\sim 600$km s$^{-1}$, and so redshift space distortions have only a small effect on the cluster correlation functions on separations greater than $\gtrsim 5$ $h^{-1}$Mpc (cf. Figures 2 and 6). At smaller separations, redshift-space distortions lead to a reduced amplitude for $\xi_{cc}$, which agrees better with the observations than the real-space correlation functions plotted in Figure 2. The APM cluster redshifts have errors of $\sim 500$km s$^{-1}$ which have not been modelled in the simulated redshift surveys. The amplitudes of the correlation functions would have been reduced slightly further on small scales had we included redshift errors in the simulated catalogues.

**Table 2.** Errors on $\xi_{cc}$

| $s$ ($h^{-1}$Mpc) | \multicolumn{3}{c}{$(\delta\xi_{cc})_P/\sigma(\xi_{cc})$} | | |
|---|---|---|---|
| | A | B | C |
| 1.33 | 0.99 | 0.90 | 0.57 |
| 2.37 | 0.65 | 0.71 | 1.20 |
| 4.22 | 1.75 | 0.83 | 1.07 |
| 7.50 | 1.27 | 0.83 | 0.82 |
| 13.3 | 0.85 | 0.70 | 0.99 |
| 23.7 | 1.07 | 0.82 | 1.05 |
| 42.2 | 0.94 | 0.56 | 0.55 |
| 75.0 | 0.89 | 0.64 | 0.60 |

The error bars on the APM data points are 'Poisson' errors computed from the formula

$$\delta\xi_{cc}(s) = \frac{(1 + \xi_{cc}(s))}{\sqrt{N_{cc}(s)}}, \qquad (10)$$

where $N_{cc}$ is the number of distinct cluster pairs in the bin centred at separation $s$ in redshift space. Our simulations show that equation (10) usually underestimates the errors on $\xi_{cc}$. In Table 2 we give the mean error on $\xi_{cc}$ for a single simulation estimated from equation (10) (($\delta\xi_{cc})_P$) divided by the scatter measured from the 10 simulations ($\sigma(\xi_{cc})$) as a function of radius for the three ensembles. Equation (10) underestimates the errors in ensemble A by a factor of about 1.1, on scales $s \gtrsim 10$ $h^{-1}$Mpc. For ensembles B and C, equation (10) underestimates the errors by a factor of about $1.5 - 1.7$ for $s \gtrsim 10$ $h^{-1}$Mpc. This is in rough agreement with the error analyses of Ling, Frenk & Barrow (1986) and Mo, Jing & Börner (1992) based on bootstrap resampling applied to real samples. Although the errors are underestimated by equation (10), none of our simulated APM surveys with $\Gamma = 0.5$ match the high amplitude of $\xi_{cc}$ for the real survey at $s \gtrsim 8$ $h^{-1}$Mpc. Thus the standard CDM model gives a poor description of the APM survey.

## 6 CONCLUSIONS

We have described a series of $N$-body simulations of CDM-like models designed to study the clustering of rich clusters. We find the following results:

[1] The cluster correlation function is insensitive to the amplitude of the mass fluctuations.

[2] The cluster correlation functions for catalogues constructed with two values of the cluster radius $r_c$, but with the same intercluster separation, are almost identical. The correlation functions are thus insensitive to the cluster selection, for values of $r_c$ comparable to those used in real cluster samples.

[3] In each set of simulations the amplitude of $\xi_{cc}$ is found to depend weakly on cluster richness. For clusters with $d_c \gtrsim 30$ $h^{-1}$Mpc, we find correlation lengths of $r_0 \approx 9$–$10$ $h^{-1}$Mpc for the standard CDM model, ($\Gamma = \Omega h = 0.5$) and $r_0 \approx 14$–$15$ $h^{-1}$Mpc for low density CDM models with $\Gamma = 0.2$.

[4] The cluster correlation functions in the low density $\Gamma = 0.2$ models are almost identical to those in the $\Omega = 1, \Gamma = 0.2$ models. The cluster correlation function in spatially flat models is therefore insensitive to the value of the cosmological density parameter but strongly dependent on the shape of the initial fluctuation spectrum.

[5] The high-peak model for $\xi_{cc}$ described in Section 4 overestimates the amplitudes of the correlation functions measured in our models, but qualitatively reproduces the observed trends with richness and with the shape of the power spectrum.

Comparing the results from our simulations with observations of the clustering of rich clusters we conclude:

[6] The standard CDM model provides a poor match to the clustering measured in the APM and other cluster redshift surveys. The APM survey is well described by a low density CDM model with $\Gamma \approx 0.2$.

[7] The weak dependence of $r_0$ on $d_c$ seen in our models does not agree with results for the clustering of rich Abell clusters. In particular, the very high amplitude $r_0 = 45 \pm 5$ $h^{-1}$Mpc measured by Peacock and West (1992) for Abell $R \geq 2$ clusters is difficult to reconcile with the CDM-like models considered here. A similar conclusion has been reached by Mann et al. (1993).

The observed clustering of galaxies agrees well with the power-spectrum of a low density, $\Gamma \approx 0.2$, CDM model (see Efstathiou, Sutherland & Maddox 1990, Vogeley et al. 1992, Loveday et al. 1992). Our results suggest that the clustering of rich clusters selected from digitised photographic plates is also consistent with such a model. However, there is potentially a discrepancy with the clustering of richer clusters selected from the Abell catalogue. If, as Bahcall and West (1992) and other authors have argued, rich clusters follow a relation of the form $r_0 = 0.4 d_c$, then we would probably have to abandon theoretical models for the origin of structure based on Gaussian initial conditions. However, the clustering of rich Abell clusters may be high because of incompletness in the Abell catalogue, and as Peacock and West note, their quoted error on $r_0$ for $R \geq 2$ could be an underestimate. Clearly it would be interesting to complete a redshift survey of several hundred rich clusters from one of the automated cluster catalogues. According to our numerical simulations, and the observational results already obtained for the APM and EDCC clusters, we would not expect $r_0$ to rise much above $\approx 16\ h^{-1}$ Mpc.

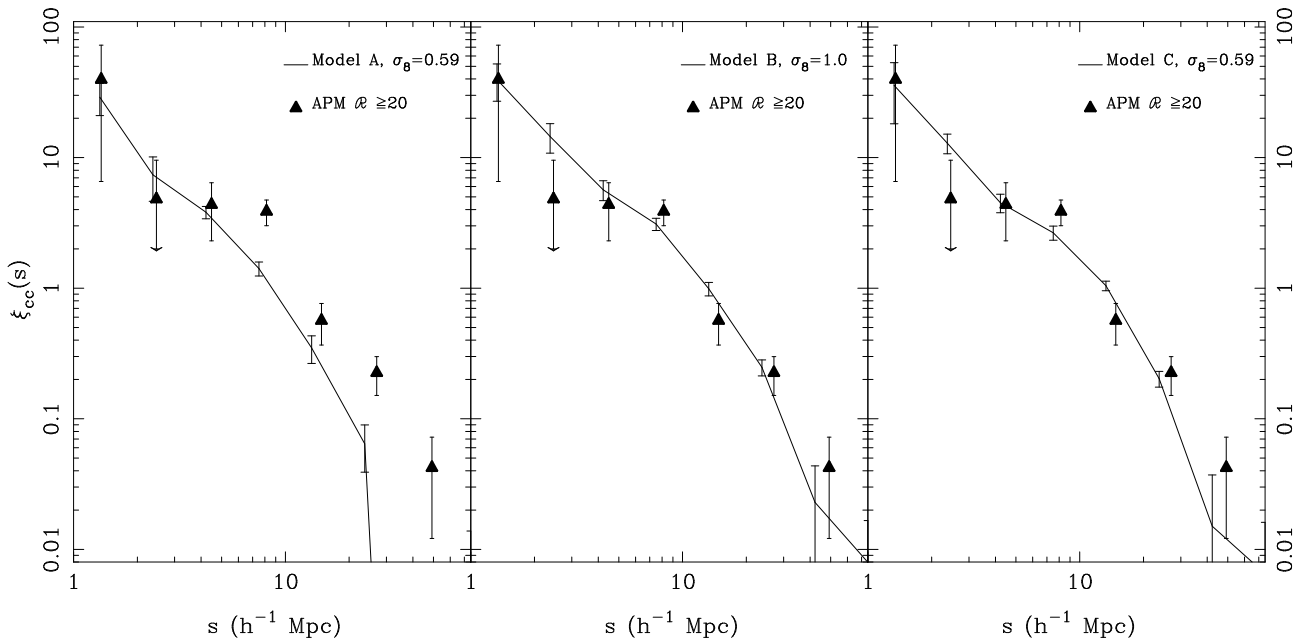

**Figure 8.** The correlation function for simulated catalogues of clusters compared to the APM data points. We plot the mean of 10 simulated catalogues in each panel and show the $1\sigma$ scatter of the mean. The errors on the APM points are the Poissonian errors computed from equation (10).

### Acknowledgments

We thank Gavin Dalton for help with various parts of this paper. We are grateful to the referee, Dr Blane Little for his very detailed report on this paper. This work was supported by grants from the UK Science and Engineering Research Council. RACC acknowledges the receipt of a SERC studentship.

### References

Abell, G. O., Corwin, H. G. & Olowin, R. P., 1989. *Ap. J. Suppl.*, **70**, 1.

Abell, G. O., 1958. *Ap. J. Suppl.*, **3**, 211.

Bahcall, N. A. & Burgett, W. S., 1986. *Ap. J. Lett.*, **300**, L35.

Bahcall, N. A. & Cen, R., 1992. *Ap. J. Lett.*, **398**, L81 (**BC92**).

Bahcall, N. A. & Soneira, R. M., 1983. *Ap. J.*, **270**, 20.

Bahcall, N. A. & West, M., 1992. *Ap. J.*, **392**, 419.

Bardeen, J. M., Bond, J. R. & Efstathiou, G., 1987. *Ap. J.*, **321**, 28.

Bardeen, J. M., Bond, J. R., Kaiser, N. & Szalay, A. S., 1986. *Ap. J.*, **304**, 15 (**BBKS**).

Bogart, R. S. & Wagoner, R. V., 1973. *Ap. J.*, **181**, 609.

Bond, J. R., Cole, S., Efstathiou, G. & Kaiser, N., 1991. *Ap. J.*, **379**, 440.

Bond, J. R. & Couchman, H. M. P., 1988. In: *Proc. Second Canadian Conference on General Relativity and Relativistic Astrophysics*, eds Coley, A., Dyer, C. & Tupper, B., World Scientific.

Bond, J. R. & Efstathiou, G., 1984. *Ap. J. Lett.*, **285**, L45.

Bower, R. G., Coles, P., Frenk, C. & White, S., 1993. *Ap. J.*, **405**, 403.

Couchman, H. M. P. & Carlberg, R. G., 1992. *Ap. J. Lett.*, **389**, 453.

Dalton, G. B., Efstathiou, G., Maddox, S. J. & Sutherland, W. J., 1992. *Ap. J. Lett.*, **390**, L1 (**DEMS**).

Davis, M., Efstathiou, G., Frenk, C. S. & White, S. D. M., 1985. *Ap. J.*, **292**, 371.

Efstathiou, G., 1993. In: *Proceedings of the U.S. National Academy of Sciences, in press*.

Efstathiou, G., Dalton, G. B., Sutherland, W. J. & Maddox, S. J., 1992. *Mon. Not. R. astr. Soc.*, **257**, 125 (**EDSM**).

Efstathiou, G., Davis, M., Frenk, C. S. & White, S. D. M., 1985. *Ap. J. Suppl.*, **57**, 241.

Efstathiou, G. & Eastwood, J. W., 1981. *Mon. Not. R. astr. Soc.*, **194**, 503.

Efstathiou, G., Sutherland, W. J. & Maddox, S. J., 1990. *Nature*, **348**, 705.

Hauser, M. G. & Peebles, P. J. E., 1973. *Ap. J.*, **185**, 757.

Holtzman, J. A. & Primack, J. R., 1993. *Ap. J.*, **405**, 428.

Kaiser, N., 1984. *Ap. J. Lett.*, **284**, L9.

Kiang, T. & Saslaw, W. C., 1969. *Mon. Not. R. astr. Soc.*, **143**, 129.

Klypin, A. A. & Kopylov, A. I., 1983. *Sov. Astron. Lett.*, **9**, 41.

Ling, E. N., Frenk, C. S. & Barrow, J. D., 1986. *Mon. Not. R. astr. Soc.*, **223**, 21p.

Loveday, J., Efstathiou, G., Peterson, B. A. & Maddox, S. J., 1992. *Ap. J. Lett.*, **400**, L43.

Mann, R. G., Heavens, A. F. & Peacock, J. A., 1993. *Mon. Not. R. astr. Soc., in press*.

Mo, H. J., Jing, Y. P. & Börner, G., 1992. *Ap. J.*, **392**, 452.

Nichol, R. C., Collins, C. A., Guzzo, L. & Lumsden, S. L., 1992. *Mon. Not. R. astr. Soc.*, **255**, 21P.

Peacock, J. A. & West, M. J., 1992. *Mon. Not. R. astr. Soc.*, **259**, 494.

Postman, M., Huchra, J. P. & Geller, M. J., 1992. *Ap. J.*, **384**, 404 (**PHG**).

Soltan, A., 1988. *Mon. Not. R. astr. Soc.*, **231**, 309.

Sutherland, W. J., 1988. *Mon. Not. R. astr. Soc.*, **234**, 159.

Sutherland, W. J. & Efstathiou, G., 1991. *Mon. Not. R. astr. Soc.*, **248**, 159.

Vogeley, M. S., Park, C., Geller, M. J. & Huchra, J. P., 1992. *Ap. J. Lett.*, **391**, L5.

White, S. D. M., Efstathiou, G. & Frenk, C., 1993. *Mon. Not. R. astr. Soc.*, **262**, 1023.

White, S. D. M., Frenk, C. S., Davis, M. & Efstathiou, G., 1987. *Ap. J.*, **313**, 505.

Yu, J. T. & Peebles, P. J. E., 1969. *Ap. J.*, **158**, 103.

Zeldovich, Ya. B., 1970. *Astron. Astrophys.*, **5**, 84.